\documentstyle[referee]{laa}

\begin{document}
\thesaurus{07  
	   (07.09.1; 
	    07.13.1;  
	   )}

\title{On Physical Interpretation of the Poynting-Robertson Effect \\
\vspace*{1.cm}
{\it Letter to the Editor}}
\author{J.~Kla\v{c}ka}
\institute{Institute of Astronomy,
   Faculty for Mathematics and Physics, Comenius University \\
   Mlynsk\'{a} dolina, 842~48 Bratislava, Slovak Republic}
\date{}
\maketitle

\begin{abstract}
Comments to the statements of Srikanth (1999) are presented.
As for the standard definition of the Poynting-Robertson drag,
the results of Srikanth (1999) are incorrect.
Srikanth's statements about the isothermality condition and ``red-shift''
are also discussed. Srikanth's results are generalized for the most general
case of the equation of motion when momentum loss per unit time
is proportional to $- \vec{v}/c$ (higher orders are neglected)
, where $\vec{v}$ is (heliocentric) velocity
of the particle, $c$ is the speed of light.


\end{abstract}

\section{Introduction}
Srikanth (1999) offers three physical viewpoints on the corresponding
statements presented in astronomical literature which is the most referenced
on the Poynting-Robertson effect (P-R effect).
It is surprising that
astronomers are still not aware of serious physical errors in paper which
they use as the most referenced literature for more than two decades.

The aim of this contribution is to discuss statements of Srikanth (1999)
in a more detail. Main attention is devoted to the
Poynting-Robertson drag (P-R drag). However, other two physical viewpoints
of Srikanth (1999) are also discussed. The discussion is presented also
for the generalization of the P-R effect, not only for perfect absorption
and symmetric reemission in particle's proper frame of reference as it
is presented in Srikanth (1999).

\section{Poynting-Robertson Drag}
The most referenced paper on the P-R effect during the last twenty years
is an invited review paper by Burns {\it et al} (1979). The definition
of the P-R drag, presented in Burns {\it et al} (1979) on page 6 states:
``The momentum loss per unit time represents ... the
{\it Poynting-Robertson drag}.''.

Let us define
\begin{equation}\label{1}
\frac{D p^{\mu}}{D \tau} \equiv
		\frac{D \left ( m~ u^{\mu} \right )}{D \tau} =
		\epsilon ~ l^{\mu} ~-~ \frac{\xi}{c^{2}} ~u^{\mu} ~,
\end{equation}
where the notation of Srikanth (1999) is used
($l^{\mu} = (1, \hat{\vec{S}}$) $\equiv$ ($1, \hat{\vec{r}}$) is not
four-vector). If we take into account the above presented definition of the
P-R drag, then the P-R drag term is given as
\begin{equation}\label{2}
\left ( \frac{D p^{\mu}}{D \tau} \right ) _{P-R ~drag} =
				   -~ \frac{\xi}{c^{2}} ~u^{\mu} ~.
\end{equation}
Thus, the statement that ``dust reemission is a necessary condition
for P-R drag as seen in the heliocentric frame'' is correct.

However, Srikanth (1999) defines P-R drag in his own way. Srikanth
rewrites Eq. (1) to the form
\begin{equation}\label{3}
	      m~  \frac{D u^{\mu}}{D \tau} =
		\epsilon ~ \left ( l^{\mu} ~-~
		\frac{l_{\nu}~u^{\nu}}{c^{2}} ~u^{\mu} \right )
\end{equation}
and he defines ``the second term on the right-hand side is the drag term''.
(One must be careful since Eq. (1) yields $\xi = \epsilon ~ l_{\nu} ~ u^{\nu}$
-- this is the reason of misunderstanding of the situation by Srikanth and
other people.)
Using this Srikanth's definition, Srikanth comes to
partially correct statement
that ``the reemission possesses an assymmetry in the heliocentric frame,
but this produces no drag'' -- there is a problem with mass $m$, since
$m$ increases due to the incident radiation. Since the P-R drag (if we want
to use such a term) should be a part of the P-R effect which yields Eq. (3)
with constant $m$, we see that the incident radiation is not able to
completely explain the P-R drag. Thus, the standard definition presented above
is more convenient and more physical.

\subsection{Generalized P-R Effect}
We want to generalize the P-R effect (Robertson 1937)
in the way that the final equation
of motion contains only two terms: the first proportional to unit
radius vector $\hat{\vec{S}} \equiv \hat{\vec{r}}$ and the second to
$- \vec{v}/c$ (higher orders are neglected),
where $\vec{v}$ is (heliocentric) velocity
of the particle and $c$ is the speed of light. The most general P-R effect
corresponds to the case when the total momentum per unit time of the
``outgoing'' radiation $\vec{p'_{o}}$ is proportional to the ``incident''
momentum per unit time $\vec{p'_{i}}$:
\begin{equation}\label{4}
\vec{p'_{o}} =  \left ( 1 ~-~ Q'_{PR} \right ) ~ \vec{p'_{i}} ~,
\end{equation}
where the primes denote quantities measured in the proper frame of
reference of the particle (see Eq. (122) in Kla\v{c}ka (1992a)).
On the basis of Eq. (4) one comes to the following equations
(special relativity is used)
\begin{eqnarray}\label{5}
\frac{d E_{p}}{d \tau} &=& Q'_{PR} ~A'~ \left ( 1 ~ - \gamma ~ w \right ) ~
			   w~ U~ c ~,
\nonumber \\
\frac{d \vec{p}}{d \tau} &=& Q'_{PR} ~ A'~
      \left ( \hat{\vec{S}} ~ - \gamma ~ w~ \frac{\vec{v}}{c} \right ) ~ w~ U~,
\end{eqnarray}
where the notation of Kla\v{c}ka (1992a) is used -- see Eqs. (133)-(134)
in Kla\v{c}ka (1992a); as for the coincidence with notation used in
Srikanth (1999), we have $l_{\mu} ~ u^{\nu} = w ~c$, $A' = \sigma$ --
cross section of the dust), $U$ -- energy density,
$u^{\mu} =$ ( $\gamma c, \gamma \vec{v}$ ), $\epsilon = A' ~U ~w$
(the right-hand side of Srikanth's Eq. (2.7) must be divided by $c$).
Eqs. (5) can be rewritten into a more compact form:
\begin{equation}\label{6}
\frac{D p^{\mu}}{D \tau} = Q'_{PR} ~A'~ w ~ U \left (
			   l^{\mu} ~-~w ~ \frac{u^{\mu}}{c} \right ) ~.
\end{equation}

The standard definition of the P-R drag (``the momentum loss per unit time'')
yields
\begin{equation}\label{7}
\left ( \frac{D p^{\mu}}{D \tau} \right ) _{P-R ~drag} =
		   -~ Q'_{PR} ~ A' ~w^{2} ~ U~  \frac{u^{\mu}}{c} \equiv
		   - \frac{\xi}{c^{2}} ~u^{\mu} ~,
\end{equation}
if the definition $\xi/c^{2}$ $\equiv$ $Q'_{PR} ~ A' ~w^{2} ~ U ~/~ c$
is used.

Using definitions $\epsilon \equiv A'~U ~w$,
$\xi/c^{2}$ $\equiv$ $Q'_{PR} ~ A' ~w^{2} ~ U ~/~ c$, we can rewrite Eq. (6)
into the form
\begin{equation}\label{8}
\frac{D p^{\mu}}{D \tau} \equiv
		\frac{D \left ( m~ u^{\mu} \right )}{D \tau} =
		\epsilon ~ l^{\mu} ~-~ \epsilon \left ( 1 ~-~ Q'_{PR} \right ) ~
		l^{\mu} ~-~ \left ( \xi / c^{2} \right ) ~ u^{\mu} ~.
\end{equation}
On the basis of $u_{\mu} ~u^{\mu} = c^{2}$, $u_{\mu} ~D~u^{\mu}/ D ~\tau = 0$,
we can rewrite Eq. (8) as
\begin{eqnarray}\label{9}
c^{2} ~ \frac{d m}{d \tau} &=& \epsilon ~ l^{\nu} ~ u_{\nu} ~-~
		\epsilon ~ \left ( 1 ~-~ Q'_{PR} \right ) ~l^{\nu} ~ u_{\nu}
		~-~ \xi ~,
\nonumber \\
m~ \frac{D u^{\mu}}{D \tau} &=&
		\epsilon ~ \left ( l^{\mu} ~-~
		\frac{l_{\nu}~u^{\nu}}{c^{2}} ~u^{\mu} \right ) ~-~
		\epsilon ~ \left ( 1 ~-~ Q'_{PR} \right ) ~
		\left ( l^{\mu} ~-~
		\frac{l_{\nu}~u^{\nu}}{c^{2}} ~u^{\mu} \right ) ~.
\end{eqnarray}
Eq. (8) is generalization of Eqs. (2.3) and (2.4) in
Srikanth (1999). We see that the interaction of the electromagnetic radiation
with the particle is important in understanding the P-R drag: constant
mass $m$ and $Q'_{PR}$ coefficient.

If we consider Eqs. (127)-(128) in Kla\v{c}ka (1992a), we can easily obtain
(special relativistic form)
\begin{equation}\label{10}
m ~\frac{d \vec{v}}{d t} = - ~ \gamma ^{-1} ~ w ~ A'~
		      \left ( 1 ~-~ Q'_{PR} \right ) ~ U
		      \left ( \hat{\vec{S}} ~-~ \frac{\vec{v}}{c} \right )
\end{equation}
for the effect of the ``outgoing'' radiation. We see that
\begin{equation}\label{11}
\left ( \frac{d \vec{v}}{d t} \right ) _{out} = 0  ~
		\Longleftrightarrow ~ Q'_{PR} = 1 ~.
\end{equation}
The case represented by Eq. (11) is the case which Srikanth wanted to stress.
However, the change of momentum is nonzero (for $\vec{v} \ne 0$),
since the change of mass is nonzero (the value is independent on
$Q'_{PR}$):
\begin{equation}\label{12}
\left ( \frac{d \vec{p}}{d t} \right ) _{out} \ne 0  ~, ~~ since ~~
\left ( \frac{d m}{d t} \right ) _{out} = ~-~
	      \gamma ^{-1} ~ w^{2} ~ U~ A'~/~c  ~.
\end{equation}

The P-R effect as a whole is important. Which part we call a P-R drag
is not important -- however, a better physical access seems to be in terms
of loss of momentum per unit time.

\section{Isothermality}
Eqs. (22), (25) and (26), Eqs. (60) and (77), and, Eqs. (133) and (141)
in Kla\v{c}ka (1992a) show that isothermality condition $d~m ~/~d~ \tau = 0$
implies conservation of energy only in the dust's rest frame. Lorentz
transformations (in special relativity) immediately show that energy changes
in heliocentric reference frame. One can immediately see this also from the
well-known relation for energy $E = \gamma ~ m ~c^{2}$, which yields
$dE/dt = \gamma ^{3} ~m ~ \vec{v} \cdot d\vec{v} / dt$ for $dm/dt = 0$ --
only in the rest frame of the particle is $dE/dt = 0$ ($\vec{v} = 0$).

However, the situation is similar to the law of reflection, which states:
``The angle of incidence equals the angle of reflection, and the incident
and reflected rays are in the same plane.''. It is supposed that
physicists know that this
formulation is correct in the rest frame of the reflecting surface and
the angles do not equal for observer who is moving with respect to the
reflecting surface.

Thus, the statement that isothermality condition implies that the dust
emits as much as it absorbs (for $Q'_{PR} = 1$) is not incorrect -- one
must only bear in mind that the formulation holds only in the dust's rest
frame. Physically educated man should know this.

\section{Red-shift}
Srikanth (1999) states that the factor $l^{\mu} ~ u_{\mu} ~/~c$  in Eq. (2.7)
does not represent red-shift. This statement is correct repetition
of the detailed discussion in Kla\v{c}ka (1992a) -- part. 2.4,
Eqs. (78)-(92), mainly Eqs. (78)-(79) and Eqs. (84)-(86).

\section{Discussion}
As we see, astronomers still do not understand physics of the P-R effect.
For the purpose of help in understanding the P-R effect author has published
papers in which the most general case of the P-R effect is derived in several
ways: Kla\v{c}ka: (1992a, 1992c, 1993a, 1993b, 1993c, 1994a).
These papers discuss
physics in detail and present physical errors in published papers, also.
Application to orbital motion is presented in
Kla\v{c}ka: (1992b -- correct statement below Eq. (22) is:
``Equations (8)-(9) and (11) still hold. Transformation $\mu \rightarrow$
$\mu ~( 1 ~-~ \beta )$ must be done in Eq. (10), now.'', 1994b, 1999),
Kla\v{c}ka and Kaufmannov\'{a} (1992).
Since even the most general case of the P-R effect still represents a very
special form of interaction between electromagnetic radiation and dust
particle, other forms of equation of motion must exist.
As for other forms of equation
of motion for dust particle due to interaction with electromagnetic radiation
we refer to Kla\v{c}ka (1993d, 1993e, 1994c, 1994d, 2000),
Kla\v{c}ka and Kocifaj (1994), Kocifaj and Kla\v{c}ka (1999).

\section{Conclusion}
We have presented several facts which should make physics of the
P-R effect more clear.

We have shown that standard definition of the P-R drag yields result which
is not consistent with the statement presented in Srikanth (1999).
However, physics yields P-R effect as a whole, and, thus, it is not wise
to separate the P-R effect into several parts; in principle it is posssible,
but different definitions may generate useless complications.

The statement that isothermality condition implies that the dust
emits as much as it absorbs (for $Q'_{PR} = 1$) is acceptable. We
have shown in an elementary manner that it automatically holds only
in the dust's rest frame. The situation is analogous to the law of reflection
-- it holds also in the rest frame of the reflecting surface.

The problem of the ``red-shift'' discussed in
Srikanth (1999) is correct repetition of the detail discussion presented
in Kla\v{c}ka (1992a).

We have generalized equations presented in
Srikanth (1999) -- Srikanth's equations hold for $Q'_{PR} = 1$.
Interesting results are given by Eqs. (10)-(12).

We have presented several references on the literature which deals with
the problematics from the physical point of view.

\acknowledgements
Special thanks to the firm ``Pr\'{\i}strojov\'{a} technika, spol. s r. o.''.
The paper was partially
supported by the Scientific Grant Agency VEGA (grant No. 1/7067/20).

\end{document}